\documentclass[12pt]{article}%
\usepackage{amsmath}
\usepackage{amsfonts}
\usepackage{amssymb}
\usepackage{graphicx}%
\providecommand{\U}[1]{\protect\rule{.1in}{.1in}}
\begin{document}
\begin{titlepage}
\ \\
\begin{center}
\LARGE
{\bf
A Brief Introduction to\\
Quantum Energy Teleportation
}
\end{center}
\ \\
\begin{center}
\large{
Masahiro Hotta
}\\
\ \\
\ \\
{\it
Department of Physics, Faculty of Science, Tohoku University,\\
Sendai 980-8578, Japan\\
hotta@tuhep.phys.tohoku.ac.jp
}
\ \\
\ \\
For invitation to SciFoo 2010\\
at Google, Mountain View, CA, USA,\\
(Jul. 30 - Aug. 1, 2010).
\end{center}
\begin{abstract}
A new quantum protocol is introduced which attains energy transportation
only by local operations and classical communication
retaining all physical laws including local energy conservation.
\end{abstract}
\end{titlepage}

\bigskip

\section{Introduction}

\ \ 

Energy transportation is a basic process for not only applied technology but
also fundamental physics. Usual methods of the transportation require physical
carriers of energy like electric currents and photons. For example, let us
consider a typical energy transportation channel like an electric cable and an
optical fiber. For the transportation, we must first infuse energy to a gate
edge A of the channel and excite energy carriers. Eventually, the energy
carriers move to an outlet edge B of the channel by time evolution of the
channel dynamics. After the carriers arrive at B, we can extract energy from
the carriers and harness it for any purpose. If the channel is in the ground
state, no activated energy carriers exist around B. Hence, by using the usual
methods, we cannot extract energy from B in the ground state.

This ground-state aspect of the usual transportation essentially remains
unchanged even if quantum effect is taken account of. In quantum theory, we
have nonvanishing zero-point energy of quantum fluctuation even in the ground
state. However, as well known, this zero-point energy at B cannot be extracted
by local operations at B. Inversely the local operations excite the quantum
fluctuation by infusing energy into the channel.

Amazingly, the situation drastically changes by adopting local operations and
classical communication of new quantum protocols called quantum energy
teleportation (QET for short) \cite{1}-\cite{8}. If we locally measure quantum
fluctuation at A in the ground state and announce the measurement result to B
at a speed much faster than the velocity of energy carriers, a part of the
zero-point energy at B can be extracted by a local operation dependent on the
measurement result before the arrival of energy carriers. \ 

This QET protocol retains all physical laws including local energy
conservation. By emitting positive energy $+E_{B}$ to outside systems, the
zero-point fluctuation at B of the channel can be more suppressed than that of
the ground state, yielding negative energy $-E_{B}$ at B. Here we fix the
origin of the energy density of the channel such that the expectational value
vanishes for the ground state. Thus the total energy of the channel is
nonnegative. In general, quantum interference among total energy eigenstates
can produce various states containing regions of such negative energy density,
although the total energy remains nonnegative.

The above local measurement at A changes the quantum state. The
post-measurement state of the channel is not the ground state but instead an
excited state with positive energy $E_{A}$. Therefore the same amount of
energy $E_{A}$ must be infused into A by the measurement device, respecting
local energy conservation law. This energy is regarded as energy input of the
teleportation. Meanwhile, the extracted energy $E_{B}$ from B is regarded as
energy output of the teleportation.

The root of the protocol is a correlation between the measurement information
at A and the quantum fluctuation at B via the ground-state entanglement. Due
to the correlation, we are able to estimate the quantum fluctuation at B based
on the announced measurement result and devise a strategy to suppress the
quantum fluctuation at B. During the selected operation on quantum fluctuation
at B generating negative energy $-E_{B}$, surplus positive energy $+E_{B}$ is
transferred from B to external systems layed at the region of B. Therefore,
QET increases not the total energy at the region of B but instead the
percentage of available energy at the region of B to be harnessed for
arbitrary purposes by decreasing the zero-point energy of B.

Physical energy carriers do not play any role for the energy extraction during
this short-time QET process. Soon after a one-round completion of the
protocol, the input energy $E_{A}$ still exists at A because late-time
evolution of the energy carries does not begin yet. Let us imagine that we
attempt to completely withdraw $E_{A}~$by local operations at A after the
extraction of energy from B. If this was possible, the energy gain $E_{B}$
might have no cost. However, if so, the total energy of the channel became
equal to $-E_{B}$ and negative. Meanwhile, we know that the total energy of
the system is nonnegative by our definition of the origin of the energy
density. Hence, it is not allowed physically to withdraw energy larger than
$E_{A}-E_{B}$ by local operations at A. This argument also implies that
$E_{A}$ is lower bounded by $E_{B}$. Another reason for this inability of
complete extraction of $E_{A}$ is because the first measurement made at A
breaks the ground-state entanglement between quantum fluctuation at A and
quantum fluctuation at B. Therefore, after the measurement at A, the ground
state (zero-energy state) is no longer recovered only by A's local operations,
which do not restore the above broken entanglement. Hence it can be concluded
that a part of input energy $E_{A}$ cannot be extracted from A during the
short time scale. QET enables this residual energy at A to be effectively
extracted in part as $E_{B}$ from the distant point B by use of the
measurement information of A. It seems like, treating the input energy $E_{A}$
as a "pawn", the quantum system "pays" the output energy $+E_{B}$ by doing
bookkeeping with a record of negative value of energy, $-E_{B}$. Needless to
say, we can harness the extracted energy $+E_{B}$ freely and do not need to
return it to the quantum system. After the completion of the QET process, a
part of the positive energy $E_{A}$ at A compensates for the negative energy
$-E_{B}$ at B during the late-time evolution of the energy-carrier dynamics.

Another way of saying QET is possible. The zero-point energy at B in the
ground state is not accessible by local operations. This looks like the energy
is saved in a locked safe under ground. In QET, we get information about the
key to open the safe by a remote measurement at A via entanglement. However,
we must then pay for it to A. The cost is energy $E_{A}$, which is larger than
the zero-point energy $E_{B}$ extracted from the safe at B.

It is worth noting that, in QET, energy can be also extracted simultaneously
from other subsystems C, D, $\cdots$ if we know the measurement result of A.
Therefore, more strictly speaking, $E_{A}$ is lower bounded by sum of all of
the possible energy extraction, $E_{B}+E_{C}+E_{D}+\cdots$. In effect, the
input energy $E_{A}$ is stored in the quantum system with a form like
broadened oil field \cite{1}.

The QET protocols can be implemented, at least theoretically, to various
physical systems, including spin chains \cite{1}-\cite{2}, cold trapped ions
\cite{3}, quantum fields \cite{4}-\cite{6} and linear harmonic chains
\cite{7}.\ Recently, a nontrivial QET protocol has been proposed for a minimal
model \cite{8}. In this presentation, analysis of the minimal QET protocol is given.

\section{Minimal QET Model}

The minimal model \cite{8} is defined as follows. The system consists of two
qubits A and B. Its Hamiltonian reads $H=H_{A}+H_{B}+V$, where each
contribution is given by
\begin{align}
H_{A}  &  =h\sigma_{A}^{z}+\frac{h^{2}}{\sqrt{h^{2}+k^{2}}},\label{1}\\
H_{B}  &  =h\sigma_{B}^{z}+\frac{h^{2}}{\sqrt{h^{2}+k^{2}}},\\
V  &  =2k\sigma_{A}^{x}\sigma_{B}^{x}+\frac{2k^{2}}{\sqrt{h^{2}+k^{2}}},
\label{3}%
\end{align}
and $h$ and$~k$ are positive constants with energy dimensions, $\sigma_{A}%
^{x}~\left(  \sigma_{B}^{x}\right)  $ is the x-component of the Pauli
operators for the qubit A (B), and $\sigma_{A}^{z}~\left(  \sigma_{B}%
^{z}\right)  $ is the z-component for the qubit A (B). The constant terms in
Eqs. (\ref{1})-(\ref{3}) are added in order to make the expectational value of
each operator zero for the ground state $|g\rangle$: $\langle g|H_{A}%
|g\rangle=\langle g|H_{B}|g\rangle=\langle g|V|g\rangle=0$. Because the lowest
eigenvalue of the total Hamiltonian $H$ is zero, $H$ is a nonnegative
operator:$~H\geq0$. Meanwhile, it should be noticed that $H_{B}$ and $H_{B}+V$
have negative eigenvalues, which can yield negative energy density at B. The
ground state is given by
\begin{align*}
|g\rangle &  =\frac{1}{\sqrt{2}}\sqrt{1-\frac{h}{\sqrt{h^{2}+k^{2}}}}%
|+\rangle_{A}|+\rangle_{B}\\
&  -\frac{1}{\sqrt{2}}\sqrt{1+\frac{h}{\sqrt{h^{2}+k^{2}}}}|-\rangle
_{A}|-\rangle_{B},
\end{align*}
where $|\pm\rangle_{A}~\left(  |\pm\rangle_{B}\right)  $ is the eigenstate of
$\sigma_{A}^{z}~\left(  \sigma_{B}^{z}\right)  $ with eigenvalue $\pm1$. A QET
protocol is constructed by the following three steps:

\begin{itemize}
\item I. A projective measurement of observable $\sigma_{A}^{x}$ is performed
to A in the ground state $|g\rangle$ and a measurement result $\left(
-1\right)  ^{\mu}$ with $\mu=0,1$ is obtained. During the measurement,
positive amount of energy%
\begin{equation}
E_{A}=\frac{h^{2}}{\sqrt{h^{2}+k^{2}}} \label{1000}%
\end{equation}
is infused to A on average.

\item II. The result $\mu$ is announced to B via a classical channel at a
speed much faster than the velocity of energy diffusion of the system.

\item III. Let us consider a local unitary operation of \thinspace B depending
on the value of $\mu$ given by $U_{B}(\mu)=I_{B}\cos\theta+i\left(  -1\right)
^{\mu}\sigma_{B}^{y}\sin\theta$, where $\theta$ is a real constant which
satisfies%
\begin{align*}
\cos\left(  2\theta\right)   &  =\frac{h^{2}+2k^{2}}{\sqrt{\left(
h^{2}+2k^{2}\right)  ^{2}+h^{2}k^{2}}},\\
\sin\left(  2\theta\right)   &  =-\frac{hk}{\sqrt{\left(  h^{2}+2k^{2}\right)
^{2}+h^{2}k^{2}}}.
\end{align*}
$U_{B}(\mu)$ is performed on B. ~During the operation, positive amount of
energy
\begin{equation}
E_{B}=\frac{h^{2}+2k^{2}}{\sqrt{h^{2}+k^{2}}}\left[  \sqrt{1+\frac{h^{2}k^{2}%
}{\left(  h^{2}+2k^{2}\right)  ^{2}}}-1\right]  \label{1001}%
\end{equation}
$~$ is extracted from $B$ on average.
\end{itemize}

\bigskip

The outline of derivation of $E_{A}$ and $E_{B}$ is given in Appendix I. The
nontrivial feature of this model is that the measurement performed at A does
not increase the average energy of B at all. By explicit calculations using
$\left[  \sigma_{A}^{x},H_{B}\right]  =\left[  \sigma_{A}^{x},V\right]  =0$,
the average values of $H_{B}$ and $V$ are found to remain zero after the
measurement and are the same as those of the ground state. Thus, we cannot
extract energy from B only by local operations soon after the measurement of
A. Even though energy carriers coming from A have not arrived at B yet, the
QET protocol is able to achieve energy extraction from B. As mentioned above,
this success of energy extraction is achieved by emergence of negative energy
density at B. Finally, it is noted that decrease of ground-state entanglement
between A and B by the measurement at A has a natural connection with the
amount of energy teleported from A to B (Appendix II).

\bigskip

\textbf{Acknowledgments}

\bigskip

I would like to thank all of the organizers and campers of SciFoo Camp. This
research has been partially supported by the Global COE Program of MEXT,
Japan, and the Ministry of Education, Science, Sports and Culture, Japan,
under Grant No. 22540406.

\appendix

\section{Appendix I. Outline of Derivation of $E_{A}$ and $E_{B}$}

In this appendix, outline of derivation of Eq. (\ref{1000}) and Eq.
(\ref{1001}) is given \cite{8}. Besides, nontrivial point of the minimal QET
model is stressed.

Firstly, the projection operator corresponding to each measurement result
$(-1)^{\mu}$ of $\sigma_{A}^{x}$ is given by
\[
P_{A}(\mu)=\frac{1}{2}\left(  1+(-1)^{\mu}\sigma_{A}^{x}\right)  .
\]
The post-measurement state of the two qubits with output $\mu$ is given by
\[
|A(\mu)\rangle=\frac{1}{\sqrt{\langle g|P_{A}(\mu)|g\rangle}}P_{A}%
(\mu)|g\rangle,
\]
where $\langle g|P_{A}(\mu)|g\rangle$ is appearance probability of $\mu\ $for
the ground state. It is easy to check that the average post-measurement state
given by
\[
\sum_{\mu}\langle g|P_{A}(\mu)|g\rangle|A(\mu)\rangle\langle A(\mu)|=\sum
_{\mu}P_{A}(\mu)|g\rangle\langle g|P_{A}(\mu)
\]
has a positive expectational value $E_{A}$ of $H$, which energy distribution
is localized at A. In fact, the value defined by
\[
E_{A}=\sum_{\mu}\langle g|P_{A}(\mu)HP_{A}(\mu)|g\rangle
\]
is computed straightforwardly as
\begin{equation}
E_{A}=\sum_{\mu}\langle g|P_{A}(\mu)H_{A}P_{A}(\mu)|g\rangle=\frac{h^{2}%
}{\sqrt{h^{2}+k^{2}}}. \label{1005}%
\end{equation}
Thus Eq.(\ref{1000}) is obtained. This infused energy $E_{A}$ is regarded as
the QET energy input via the measurement of A. During the measurement, $E_{A}$
is transferred from external systems including the measurement device with a
battery respecting local energy conservation. The QET energy conservation law
during local measurements has been discussed in \cite{2}.

Because energy of B remains zero after the measurement, we cannot extract
energy from B by local operations soon after the measurement. To verify this
fact explicitly, let us consider any local unitary operation $W_{B}$ which is
\textbf{independent }of A's measurement result and performed on B.\quad Then,
the post-operation state $\omega$ is given by
\[
\omega=\sum_{\mu}W_{B}P_{A}(\mu)|g\rangle\langle g|P_{A}(\mu)W_{B}^{\dag}.
\]
The energy difference after the operation is calculated as
\begin{equation}
E_{A}-\operatorname*{Tr}\left[  \omega H\right]  =-\langle g|W_{B}^{\dag
}\left(  H_{B}+V\right)  W_{B}|g\rangle, \label{e01}%
\end{equation}
where we have used
\[
W_{B}^{\dag}H_{A}W_{B}=H_{A}W_{B}^{\dag}W_{B}=H_{A},
\]%
\[
\left[  W_{B}^{\dag}\left(  H_{B}+V\right)  W_{B},~P_{A}(\mu)\right]  =0,
\]
and the completeness relation of $P_{A}(\mu)$:%
\[
\sum_{\mu}P_{A}(\mu)=1_{A}.
\]
From Eq. (\ref{e01}), it is proven that the energy difference is nonpositive:
\[
E_{A}-\operatorname*{Tr}\left[  \omega H\right]  =-\langle g|W_{B}^{\dag
}HW_{B}|g\rangle\leq0,
\]
because of a relation such that $\langle g|W_{B}^{\dag}H_{A}W_{B}%
|g\rangle=\langle g|H_{A}|g\rangle=0$ and the nonnegativity of $H$. Therefore,
as a natural result, no local operation on B independent of $\mu$ extracts
positive energy from B by decreasing total energy of the two qubits.

After a while, the infused energy $E_{A}$ diffuses to B. The time evolution of
the expectational values $H_{B}$ and $V$ of the average post-measurement state
is calculated as%
\begin{align*}
\langle H_{B}(t)\rangle &  =\sum_{\mu}\langle g|P_{A}(\mu)|g\rangle\langle
A(\mu)|e^{itH}H_{B}e^{-itH}|A(\mu)\rangle\\
&  =\frac{h^{2}}{2\sqrt{h^{2}+k^{2}}}\left[  1-\cos\left(  4kt\right)
\right]  ,
\end{align*}
and $\langle V(t)\rangle=0$. Therefore, energy can be extracted from B after a
diffusion time scale of $1/k$; this is just a usual energy transportation from
A to B. The QET protocol can transport energy from A to B in a time scale much
shorter than that of this usual transportation. In the \ protocol, the
measurement output $\mu$ is announced to B. Because the model is
non-relativistic, the propagation speed of the announced output can be much
faster than the diffusion speed of the infused energy and can be approximated
as infinity. Soon after the arrival of the output $\mu$, we perform $U_{B}%
(\mu)$ on B dependent on $\mu$. Then, the average state after the operation is
given by%

\[
\rho=\sum_{\mu}U_{B}(\mu)P_{A}(\mu)|g\rangle\langle g|P_{A}(\mu)U_{B}%
(\mu)^{\dag}.
\]
The expectational value of the total energy after the operation is given by
\[
\operatorname*{Tr}\left[  \rho H\right]  =\sum_{\mu}\langle g|P_{A}(\mu
)U_{B}(\mu)^{\dag}HU_{B}(\mu)P_{A}(\mu)|g\rangle.
\]
On the basis of the fact that $U_{B}(\mu)$ commutes with $H_{A}$ and Eq.
(\ref{1005}), $E_{B}$ is computed as
\[
E_{B}=E_{A}-\operatorname*{Tr}\left[  \rho H\right]  =-\operatorname*{Tr}%
\left[  \rho\left(  H_{B}+V\right)  \right]  .
\]
Further, on the basis of the fact that $P_{A}(\mu)$ commutes with $U_{B}(\mu
)$, $H_{B}$ and $V$, the energy can be written as
\[
E_{B}=-\sum_{\mu}\langle g|P_{A}(\mu)\left(  H_{B}(\mu)+V(\mu)\right)
|g\rangle,
\]
where the $\mu$-dependent operators are given by $H_{B}(\mu)=U_{B}(\mu)^{\dag
}H_{B}U_{B}(\mu)$ and $V(\mu)=U_{B}(\mu)^{\dag}VU_{B}(\mu)$. By
straightforward calculation, $E_{B}$ is computed as
\begin{equation}
E_{B}=-\frac{1}{\sqrt{h^{2}+k^{2}}}\left[  \left(  h^{2}+2k^{2}\right)
\left[  1-\cos\left(  2\theta\right)  \right]  +hk\sin(2\theta)\right]  .
\label{1002}%
\end{equation}
Note that $E_{B}=0$ if $\theta=0$, as it should be. If we take a small
negative value of $\theta$ in Eq. (\ref{1002}), it is noticed that $E_{B}$
takes a small positive value such that%
\[
E_{B}\sim\frac{2hk\left\vert \theta\right\vert }{\sqrt{h^{2}+k^{2}}}>0.
\]
Maximization of $E_{B}$ in terms of $\theta$ is attained by taking a value of
$\theta$ which satisfies
\begin{align*}
\cos\left(  2\theta\right)   &  =\frac{h^{2}+2k^{2}}{\sqrt{\left(
h^{2}+2k^{2}\right)  ^{2}+h^{2}k^{2}}},\\
\sin\left(  2\theta\right)   &  =-\frac{hk}{\sqrt{\left(  h^{2}+2k^{2}\right)
^{2}+h^{2}k^{2}}}.
\end{align*}
Substituting these relations into Eq. (\ref{1002}) yields the positive value
of $E_{B}$ in Eq. (\ref{1001}).

The measurement of A can be extended to POVM measurements. Let $S_{M_{A}}$
denote a set of POVM measurements for A which measurement operators $M_{A}%
(\mu)$ with measurement output $\mu$ commute with the interaction Hamiltonian
$V$. The measurement operator $M_{A}(\mu)$ takes the form of
\begin{equation}
M_{A}(\mu)=e^{i\delta_{\mu}}\left(  m_{\mu}+e^{i\alpha_{\mu}}l_{\mu}\sigma
_{A}^{x}\right)  . \label{1008}%
\end{equation}
The coefficients $m_{\mu}$, $l_{\mu}$, $\alpha_{\mu}$ and $\delta_{\mu}$ are
real constants which satisfy
\begin{align*}
\sum_{\mu}\left(  m_{\mu}^{2}+l_{\mu}^{2}\right)   &  =1,\\
\sum_{\mu}m_{\mu}l_{\mu}\cos\alpha_{\mu}  &  =0.
\end{align*}
The POVM corresponding to $M_{A}(\mu)$ is defined by
\[
\Pi_{A}(\mu)=M_{A}(\mu)^{\dag}M_{A}(\mu),
\]
which satisfies the completeness relation,
\[
\sum_{\mu}\Pi_{A}(\mu)=1_{A}.
\]
By introducing the emergence probability $p_{A}(\mu)=\langle g|\Pi_{A}%
(\mu)|g\rangle$ of output $\mu$ for the ground state and a real parameter
$q_{A}(\mu)$, the POVM is written as follows:
\[
\Pi_{A}(\mu)=p_{A}(\mu)+q_{A}(\mu)\sigma_{A}^{x}.
\]
By taking suitable values of $m_{\mu}$, $l_{\mu}$, and $\alpha_{\mu}$, all
values of $p_{A}(\mu)$ and $q_{A}(\mu)$ are permissible as long as they
satisfy $\sum_{\mu}p_{A}(\mu)=1$, $\sum_{\mu}q_{A}(\mu)=0$ and $p_{A}(\mu
)\geq\left\vert q_{A}(\mu)\right\vert $. The post-measurement state of the two
qubits with output $\mu$ is given by
\begin{equation}
|A^{\prime}(\mu)\rangle=\frac{1}{\sqrt{p_{A}(\mu)}}M_{A}(\mu)|g\rangle.
\label{1009}%
\end{equation}
This measurement excites the system. Input energy $E_{A}$ of QET in this case
is defined by%
\[
E_{A}=\sum_{\mu}\langle g|M_{A}(\mu)^{\dag}HM_{A}(\mu)|g\rangle
\]
and is computed as
\[
E_{A}=\frac{2h^{2}}{\sqrt{h^{2}+k^{2}}}\sum_{\mu}l_{\mu}^{2}.
\]
It is also possible to generalize the operation of B as%
\begin{equation}
U_{B}^{\prime}(\mu)=I_{B}\cos\omega_{\mu}+i\vec{n}_{\mu}\cdot\vec{\sigma}%
_{B}\sin\omega_{\mu}, \label{1007}%
\end{equation}
where $\omega_{\mu}\,$\ is a real parameter, $\vec{n}_{\mu}=\left(  n_{x\mu
},n_{y\mu},n_{z\mu}\right)  \,$is a three-dimensional unit real vector and
$\vec{\sigma}_{B}$ is the Pauli spin vector operator of B. After the operation
of B, the average state becomes%

\[
\rho^{\prime}=\sum_{\mu}U_{B}^{\prime}(\mu)M_{A}(\mu)|g\rangle\langle
g|M_{A}(\mu)^{\dag}U_{B}^{\prime}(\mu)^{\dag}.
\]
Output energy $E_{B}$ of QET is defined by%
\[
E_{B}=E_{A}-\operatorname*{Tr}\left[  \rho^{\prime}H\right]
\]
and computed as
\begin{equation}
E_{B}=\frac{1}{\sqrt{h^{2}+k^{2}}}\sum_{\mu}Q(\mu), \label{1010}%
\end{equation}
where $Q(\mu)$ is given by
\[
Q(\mu)=X(\mu)\cos\left(  2\omega_{\mu}\right)  -hkq_{A}(\mu)n_{y\mu}%
\sin(2\omega_{\mu})-X(\mu),
\]
and $X(\mu)$ is defined by
\[
X(\mu)=p_{A}(\mu)\left[  h^{2}\left(  1-n_{z\mu}^{2}\right)  +2k^{2}\left(
1-n_{x\mu}^{2}\right)  \right]  -3hkq_{A}(\mu)n_{x\mu}n_{z\mu}.
\]
It can proven that, for each measurement belonging to $S_{M_{A}}$, an
operation $U_{B}^{\prime}(\mu)$ properly dependent on $M_{A}(\mu)$ and $\mu$
always yields a positive value of $E_{B}$ \cite{8}.

\section{Appendix II. Energy-Entanglement Relation for Minimal QET Model}

In this appendix, we analyze entanglement breaking by the measurement of A and
show two inequalities between entanglement consumption in the measurement and
amount of teleported energy \cite{8}. We adopt entropy of entanglement as a
quantitative measure of entanglement. The entropy of a pure state $|\Psi
_{AB}\rangle$ of $A$ and $B$ is defined as
\[
S_{AB}=-\operatorname*{Tr}_{B}\left[  \operatorname*{Tr}_{A}\left[  |\Psi
_{AB}\rangle\langle\Psi_{AB}|\right]  \ln\operatorname*{Tr}_{A}\left[
|\Psi_{AB}\rangle\langle\Psi_{AB}|\right]  \right]  .
\]
Before the measurement, the total system is prepared to be in the ground state
$|g\rangle$. \ The reduced state of B is given by
\[
\rho_{B}=\operatorname*{Tr}_{A}\left[  |g\rangle\langle g|\right]  .
\]
After the POVM measurement outputting $\mu$ defined by Eq. (\ref{1008}), the
state is transferred into a pure state $|A^{\prime}(\mu)\rangle$ in Eq.
(\ref{1009}). The reduced post-measurement state of B is calculated as
\[
\rho_{B}(\mu)=\frac{1}{p_{A}(\mu)}\operatorname*{Tr}_{A}\left[  \Pi_{A}%
(\mu)|g\rangle\langle g|\right]  .
\]
The entropy of entanglement of the ground state is given by
\[
S_{AB}(g)=-\operatorname*{Tr}_{B}\left[  \rho_{B}\ln\rho_{B}\right]
\]
and that of the post-measurement state with output $\mu$ is given by
\[
S_{AB}(\mu)=-\operatorname*{Tr}_{B}\left[  \rho_{B}(\mu)\ln\rho_{B}%
(\mu)\right]  .
\]
By using these results, we define the consumption of ground-state entanglement
by the measurement as the difference between the ground-state entanglement and
the averaged post-measurement-state entanglement:%

\[
\Delta S_{AB}=S_{AB}(g)-\sum_{\mu}p_{A}(\mu)S_{AB}(\mu).
\]
Interestingly, this quantity is tied to the quantum mutual information between
the measurement result of A and the post-measurement state of B. Let us
introduce a Hilbert space for a measurement pointer system $\bar{A}$ of the
POVM measurement, which is spanned by orthonormal states $|\mu_{\bar{A}%
}\rangle$ corresponding to the output $\mu$ satisfying $\langle\mu_{\bar{A}%
}|\mu_{\bar{A}}^{\prime}\rangle=\delta_{\mu\mu^{\prime}}$. Then, the average
state of $\bar{A}$ and B after the measurement is given by
\[
\Phi_{\bar{A}B}=\sum_{\mu}p_{A}(\mu)|\mu_{\bar{A}}\rangle\langle\mu_{\bar{A}%
}|\otimes\rho_{B}(\mu).
\]
By using the reduced operators $\Phi_{\bar{A}}=\operatorname*{Tr}_{B}\left[
\Phi_{\bar{A}B}\right]  $ and $\Phi_{B}=\operatorname*{Tr}_{\bar{A}}\left[
\Phi_{\bar{A}B}\right]  $, the mutual information $I_{\bar{A}B}$ is defined as%
\begin{equation}
I_{\bar{A}B}=-\operatorname*{Tr}_{\bar{A}}\left[  \Phi_{\bar{A}}\ln\Phi
_{\bar{A}}\right]  -\operatorname*{Tr}_{B}\left[  \Phi_{B}\ln\Phi_{B}\right]
+\operatorname*{Tr}_{\bar{A}B}\left[  \Phi_{\bar{A}B}\ln\Phi_{\bar{A}%
B}\right]  .\nonumber
\end{equation}
By using $\operatorname*{Tr}_{B}\left[  \Phi_{\bar{A}B}\right]  =\sum_{\mu
}p_{A}(\mu)|\mu_{\bar{A}}\rangle\langle\mu_{\bar{A}}|$ and $\operatorname*{Tr}%
_{\bar{A}}\left[  \Phi_{\bar{A}B}\right]  =\sum_{\mu}p_{A}(\mu)\rho_{B}%
(\mu)=\rho_{B}$, it can be straightforwardly proven that $I_{\bar{A}B}$ is
equal to $\Delta S_{AB}$. This relation provides another physical
interpretation of $\Delta S_{AB}$.

Next, let us calculate $\Delta S_{AB}$ explicitly. All the eigenvalues of
$\rho_{B}(\mu)$ are given by%

\begin{equation}
\lambda_{\pm}(\mu)=\frac{1}{2}\left[  1\pm\sqrt{\cos^{2}\varsigma+\sin
^{2}\varsigma\frac{q_{A}(\mu)^{2}}{p_{A}(\mu)^{2}}}\right]  , \label{29}%
\end{equation}
where $\varsigma$ is a real constant which satisfies
\[
\cos\varsigma=\frac{h}{\sqrt{h^{2}+k^{2}}},~\sin\varsigma=\frac{k}{\sqrt
{h^{2}+k^{2}}}.
\]
The eigenvalues of $\rho_{B}$ are obtained by substituting $q_{A}(\mu)=0$ into
Eq. (\ref{29}). By using $\lambda_{s}(\mu)$, $\Delta S_{AB}$ can be evaluated
as
\begin{equation}
\Delta S_{AB}=\sum_{\mu}p_{A}(\mu)f_{I}\left(  \frac{q_{A}(\mu)^{2}}{p_{A}%
(\mu)^{2}}\right)  , \label{30}%
\end{equation}
where $f_{I}(x)$ is a monotonically increasing function of $x\in\left[
0,1\right]  $ and is defined by%
\begin{align*}
f_{I}(x)  &  =\frac{1}{2}\left(  1+\sqrt{\cos^{2}\varsigma+x\sin^{2}\varsigma
}\right) \\
&  \times\ln\left(  \frac{1}{2}\left(  1+\sqrt{\cos^{2}\varsigma+x\sin
^{2}\varsigma}\right)  \right) \\
&  +\frac{1}{2}\left(  1-\sqrt{\cos^{2}\varsigma+x\sin^{2}\varsigma}\right) \\
&  \times\ln\left(  \frac{1}{2}\left(  1-\sqrt{\cos^{2}\varsigma+x\sin
^{2}\varsigma}\right)  \right) \\
&  -\frac{1}{2}\left(  1+\cos\varsigma\right)  \ln\left(  \frac{1}{2}\left(
1+\cos\varsigma\right)  \right) \\
&  -\frac{1}{2}\left(  1-\cos\varsigma\right)  \ln\left(  \frac{1}{2}\left(
1-\cos\varsigma\right)  \right)  .
\end{align*}
It is worth noting \cite{8} that the maximum of $E_{B}$ of Eq. (\ref{1010}) in
terms of $U_{B}^{\prime}(\mu)$ of Eq. (\ref{1007}) takes a form similar to Eq.
(\ref{30}) as%

\begin{equation}
\max_{U_{B}^{\prime}(\mu)}E_{B}=\sum_{\mu}p_{A}(\mu)f_{E}\left(  \frac
{q_{A}(\mu)^{2}}{p_{A}(\mu)^{2}}\right)  , \label{003}%
\end{equation}
where $f_{E}(x)$ is a monotonically increasing function of $x\in\left[
0,1\right]  $ and is defined by
\[
f_{E}(x)=\sqrt{h^{2}+k^{2}}\left(  1+\sin^{2}\varsigma\right)  \left[
\sqrt{1+\frac{\cos^{2}\varsigma\sin^{2}\varsigma}{\left(  1+\sin^{2}%
\varsigma\right)  ^{2}}x}-1\right]  .
\]
Expanding both $f_{I}(x)$ and $f_{E}(x)$ around $x=0$ yields%

\begin{align*}
f_{I}(x)  &  =\frac{\sin^{2}\varsigma}{4\cos\varsigma}\ln\frac{1+\cos
\varsigma}{1-\cos\varsigma}x+O(x^{2}),\\
f_{E}(x)  &  =\sqrt{h^{2}+k^{2}}\frac{\cos^{2}\varsigma\sin^{2}\varsigma
}{2\left(  1+\sin^{2}\varsigma\right)  }x+O(x^{2}).
\end{align*}
By deleting $x$ in the above two equations, we obtain the following relation
for weak measurements with infinitesimally small $q_{A}(\mu)$:%

\[
\Delta S_{AB}=\frac{1+\sin^{2}\varsigma}{2\cos^{3}\varsigma}\ln\frac
{1+\cos\varsigma}{1-\cos\varsigma}\frac{\max_{U_{B}^{\prime}(\mu)}E_{B}}%
{\sqrt{h^{2}+k^{2}}}+O(q_{A}(\mu)^{4}).
\]
It is of great significance \cite{8} that this relation can be extended as the
following inequality for general measurements of $S_{M_{A}}:$%

\begin{equation}
\Delta S_{AB}\geq\frac{1+\sin^{2}\varsigma}{2\cos^{3}\varsigma}\ln\frac
{1+\cos\varsigma}{1-\cos\varsigma}\frac{\max_{U_{B}^{\prime}(\mu)}E_{B}}%
{\sqrt{h^{2}+k^{2}}}. \label{32}%
\end{equation}
This inequality implies that a large amount of teleported energy requests a
large amount of consumption of the ground-state entanglement between A and B.
In addition, we can prove another inequality between the teleported energy and
the entanglement consumption \cite{8}. The following inequality is satisfied
for all measurements of $S_{M_{A}}$:
\begin{align}
&  \max_{U_{B}^{\prime}(\mu)}E_{B}\nonumber\\
&  \geq\frac{2\sqrt{h^{2}+k^{2}}\left[  \sqrt{4-3\cos^{2}\varsigma}-2+\cos
^{2}\varsigma\right]  }{\left(  1+\cos\varsigma\right)  \ln\left(  \frac
{2}{1+\cos\varsigma}\right)  +\left(  1-\cos\varsigma\right)  \ln\left(
\frac{2}{1-\cos\varsigma}\right)  }\nonumber\\
&  \times\Delta S_{AB}. \label{1077}%
\end{align}
This ensures that if we have consumption of ground-state entanglement $\Delta
S_{AB}$ for a measurement of $S_{M_{A}}$, we can in principle teleport energy
from A to B, where the energy amount is greater than the value of the
right-hand-side term of Eq. (\ref{1077}). This bound is achieved for non-zero
energy transfer by measurements with $q_{A}(\mu)=\pm p_{A}(\mu)$. The
inequalities in Eq. (\ref{32}) and Eq. (\ref{1077}) help us to gain a deeper
understanding of entanglement as a physical resource because they show that
the entanglement decrease by the measurement of A is directly related to the
increase of the available energy at B as an evident physical resource.

\bigskip
\end{document}